# Optimizing microlens arrays for incoherent HiLo microscopy

**Ziao Jiao[1,2], Xi Chen[1], and David Day Uei Li[1]**

[1] Department of Biomedical Engineering, University of Strathclyde, Glasgow G1 1XQ, Scotland, UK
[2] Strathclyde Institute of Pharmacy and Biomedical Sciences, University of Strathclyde, Glasgow G4 0RE, Scotland, UK

E-mail: David.li@strath.ac.uk

**Abstract**

HiLo microscopy is a powerful, low-cost, and easily configurable technique for acquiring high-contrast optically-sectioned images. However, traditional HiLo microscopes are based on coherent light sources with diffusive glass plates or incoherent light sources with digital mirror devices (DMD) and spatial light modulators (SLM), which are more expensive. Here, we propose a new low-cost HiLo microscopy technique using MLAs and incoherent LED light sources. We simulated structured illumination (SI) patterns and HiLo image generation based on Fresnel diffraction and incoherent imaging. To observe how MLAs affect HiLo images, we used three common MLAs with specific microlens pitch and numerical aperture (NA) to generate periodic illumination patterns. According to our simulations, using MLAs and incoherent light sources can enhance the image contrast compared with a traditional widefield fluorescence microscope. We found that the MLA's NA does not significantly affect HiLo images. A larger lens pitch can bring a higher image contrast. However, there is an optimized lens pitch. If the lens pitch is too high, artefacts are observed in HiLo images. To our knowledge, this is the first numerical study about MLA-based HiLo microscopy. This study can benefit researchers using MLAs and incoherent light sources to configure their low-cost HiLo microscopes.

## 1. Introduction

HiLo is a widefield fluorescence microscope with an optical-sectioning (OS) function. It is a robust method for rejecting background noise and acquiring high-resolution OS images [1,2]. HiLo can reveal fast biological processes in neuron cells [3,4], discover 3D cell mechanical properties by rejecting out-of-focus signals [5], improve retinal image quality [6], and realize 3D image cytometers [7]. HiLo can also achieve rapid and non-destructive imaging of freshly excised tissues with high efficiency and a good resolution [8]. Besides these biomedical applications, HiLo can measure engineering microstructure surface profiles [9]. HiLo's configuration is like widefield epi-fluorescence microscopy (Fig.1). The only difference is that in HiLo, the illumination light is adjusted for generating patterns on the sample plane. Any widefield epi-fluorescence microscopy can be refitted to HiLo for optically-sectioned imaging by changing the illumination path. Moreover, HiLo can be easily combined with other biomedical imaging modalities, such as endoscopy [10–12], 3D volumetric microscopy [13,14], and optical scanning microscopy [15]. Compared with OS microscopy techniques such as confocal [16], two-photon [17], and light sheet microscopy [18], HiLo does not need pinhole assignments, an expensive laser module, and dual objective lenses for illumination and excitation [19]. Similar to structured illumination (SI) microscopy, the light source in HiLo project a specific SI pattern to the in-focus sample plane. However, only two images are required for HiLo to reconstruct high-contrast OS images [20] instead of three images for the OS SI microscope [21], one acquired with widefield illumination and the other with SI. With deep learning algorithms and advanced image processing, HiLo can now realize OS imaging using only one image [22,23]. All these properties make HiLo relatively low-cost and easy to operate.

HiLo's SI pattern determines the OS image's contrast and axial resolution [24]. Several studies have concluded that we can optimize the properties of speckle illumination [25,26] and frequency, modulation depth, and SI patterns [12, 27] for the best OS images. There are two common ways to obtain SI patterns in HiLo: 1) using a coherent light source with a diffuser [28, 29] to create speckle patterns or 2) using an incoherent light source with diffraction optical devices [4, 30] to develop periodic patterns. However, coherent light sources, spatial light modulators (SLMs), and digital mirror devices (DMDs) increase the cost and system complexity. Microlens arrays (MLA) can also modulate light into different patterns. Researchers have exploited lithography [31] and laser etching techniques [32] to fabricate MLAs quickly. Moreover, the moulding method can also produce PDMS and PMMA-based MLAs [33]. Using appropriate MLAs instead of DMD or LSMs for HiLo can minimize costs and simplify the optical path.

Moreover, a cheaper LED uncoherent light source can replace the coherent light source in HiLo. Therefore, using MLAs and LEDs could significantly reduce the cost and complexity. Choosing a specific MLA with proper parameters is essential to realize this idea.

This numerical simulation study assesses the integration of uncoherent LEDs and MLAs for a low-cost HiLo system. We first simulated how MLAs generate periodic illumination patterns on the illuminated sample volume. According to several studies about MLAs fabrication [34, 35], we chose the three most common types of MLAs here: the cross-, cylinder-, and hexagon-types (Fig.1). We examined how MLA's NA and its microlens pitch affects the spatial distribution of these illumination patterns. Then we generated a simulated fluorescent block, multiplying it with the MLA-generated illumination pattern and using the HiLo algorithm to obtain high-contrast optical-sectioning images. To balance the computational efficiency and mesh fineness, we tuned MLA's NA within a small range (0.006-0.01). The microlens pitch is set from 60μm to 140μm with a 20μm step. After passing through the optical illumination path with 0.1 times magnification, the illumination pattern period is from 6μm to 14μm with a 2μm step.

We discovered from the simulation results that the HiLo image contrast is not significantly affected by MLA's NA but still has a minor enhancement with a higher NA. Noticeably, the relationship between the microlens pitch and image contrast is more prominent. Moreover, a higher period can enhance the image contrast, but if the illumination period is 14μm/period, the contrast deteriorates again. We also examined three MLA types to discover how pattern distribution affects image quality. Three kinds of MLAs perform similarly, except the illumination period reaches its optimal value (12μm/period), and the cross-type MLA performs better than the others. We can obtain the best results using the cross-type MLA with 0.1NA and a pitch of 120μm. We believe this numerical study can encourage more theoretical studies and the future development of low-cost MLA-based HiLo microscopes.

## 2. Theory

### *2.1 Fundamentals of HiLo microscope*

The mathematic deduction and theoretical model of HiLo are reported in detail else [20,25,28,29,36]. Therefore we only describe it briefly. In HiLo, two images combine an optical-sectioning image ($I_{HiLo}(u,v)$):

$$I_{HiLo}(u,v) = I_{Hi}(u,v) + \eta I_{Lo}(u,v), \tag{1}$$

where $I_{Hi}(u,v)$ is the in-focus high-frequency image, $I_{Lo}(u,v)$ is the in-focus low-frequency image, and $u,v$ are spatial coordinates. The parameter $\eta$ can avoid discontinuities in the frequency domain, which can be calculated by [12]:

$$\eta = \frac{HP_{k_c}}{LP_{k_c}}, \tag{2}$$

where $HP_{\kappa_c}$ and $LP_{\kappa_c}$ are Gaussian high-pass and low-pass filters, respectively, and $k_c$, the cut-off frequency, should be less than or equal to the frequency of the structured illumination pattern. $LP_{\kappa_c}$ is complementary to $HP_{\kappa_c}$. Because $I_{Hi}(u,v)$ is intuitively axially resolved, it can be acquired easily by:

$$I_{Hi}(u,v) = \mathcal{F}^{-1}\{HP_{k_c}[I_u(k_u,k_v)]\}, \tag{3}$$

where $I_u(k_u, k_v)$ is the captured image in the frequency domain under uniform illumination and $\mathcal{F}^{-1}\{\}$ is the inverse Fourier operator. To get $I_{Lo}(u,v)$, which cannot be axially resolved, we need a weighting function to select the in-focus portion of $I_u(u,v)$ below the cut-off frequency. First, we get a bias-free difference image $I_\Delta(u,v)$ by subtracting $I_s(u,v)$, the captured image in the spatial domain under structured illumination, from $I_u(u,v)$:

$$I_\Delta(u,v) = I_u(u,v) - I_s(u,v), \tag{4}$$

where $I_u(u,v)$ and $I_s(u,v)$ are the captured images under uniform and structured illumination in the spatial domain. According to Eq. (4), we can estimate the illumination-induced contrast $C(u,v)$ by:

$$C(u,v) = \sigma\{I_\Delta(u,v)\}, \tag{5}$$

where $\sigma\{\}$ means the standard deviation. According to Eq. (4) and Eq. (5), we can guarantee that the modulated component is locally centered about zero and make the evaluation insensitivity to differences in the global illumination profile [10]. Then we weighted $I_u(u,v)$ with $C(u,v)$ and input it to the low-pass filter to acquire $I_{Lo}(u,v)$:

$$I_{Lo}(u,v) = \mathcal{F}^{-1}\{LP_{k_c}[C(k_u,k_v)I_u(k_u,k_v)]\}, \tag{6}$$

where $C(k_u,k_v)$ is $C(u,v)$ in the frequency domain.

## 2.2 Theory of pattern projection and image formation

To study how MLAs affect the image quality of a HiLo microscope, we give a detailed mathematic deduction about structured illumination and image formation. For efficiency, we performed multiplications in the spectral domain instead of convolutions in the spatial domain. Fig. 1 illustrates the physical model of this study. An MLA modulates the collimated beam and forms a specific pattern on its focal plane. We use the Fresnel diffraction theory in this part. To create different patterns, we used three common types of MLAs (cross-, cylinder- and hexagon types). The illumination part of the 4f microscope system (L1 and OL) conjugates the illumination pattern to the sample plane, and the camera captures the excited fluorescent signal through the imaging part of the 4f microscope system (OL and Tube lens). We used the incoherent imaging theory in the illumination and imaging parts.

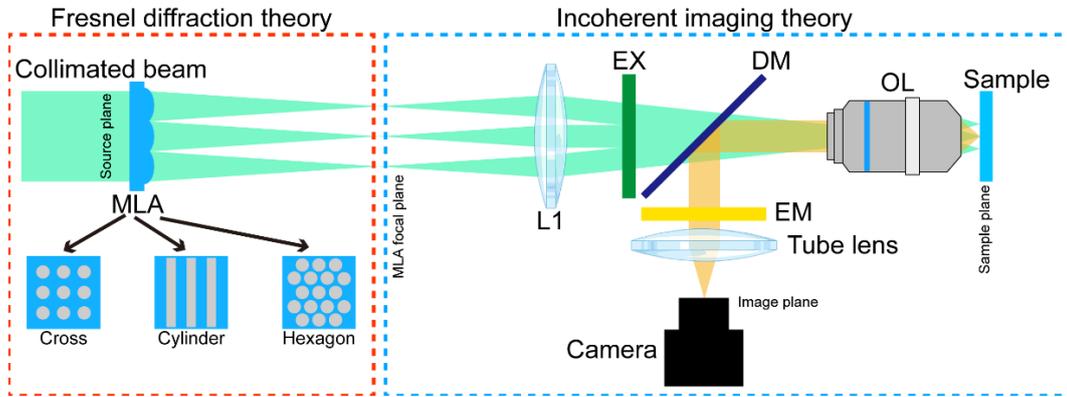

Fig. 1. The physical model of this study. Red and blue dashed rectangles denote physical theories used in different parts. We used three different MLAs to generate different illumination patterns. This optical path is for structured illumination. The uniform illumination can be easily obtained by removing the MLA, and the collimated beam can be focused on the back pupil plane of the OL by L1. For simplicity, we denote four planes in the figure (source plane, MLA focal plane, sample plane, and image plane), which are the same as Fig. 2. (MLA: microlens array; L1: convex lens; EX: excitation filter; DM: dichroic mirror; OL: objective lens; EM: emission filter).

Firstly, let us see the image formation on the image plane. Fig. 2 is the coordinate system for a mathematical deduction. The image acquired under structured illumination is:

$$I_{4s}(u,v) = \sum_{z_i=0}^{n} \iint PSF_{em}(u-\mu, v-\sigma)|_{z=z_i} I_3(\mu,\sigma,z_i) O(\mu,\sigma,z_i) \, d\mu d\sigma, \tag{7}$$

where $PSF_{em}$ is the 2D emission point spread function with a defocus term (different $z$). $O$ and $I_3$ are the 3D objects and illumination distribution, respectively, and $z_i$ is the defocus distance between the focus and target plane ($z_i = 0$ is in focus).

To obtain the intensity distribution on the image plane, we convoluted the corresponding structure-illuminated sample plane for each slice with a defocused $PSF_{em}$, and summed them up from $z_i = 0$ to $z_i = n$. When $I_3(\mu, \sigma, z_i)$ is 1 for different $z_i$, we can obtain the image under uniform illumination ($I_{4u}(u,v)$). For simplicity, we assume the imaging system is telecentric and has unit magnification. From Eq. (6), $I_3(\mu, \sigma, z_i)$ can determine the final HiLo image.

The incident beam is assumed to be a monochromatic plane wave, and we only consider the distribution inside each microlens of the MLA. Here, we consider the cross-type MLA and the amplitude distribution before the MLA can be written as:

$$U_1(\xi, \eta) = \sum_i \sum_j circ\left[\frac{\sqrt{(\xi-id)^2+(\eta-jd)^2}}{w}\right] = \begin{cases} 1 & \text{inside each micro lens} \\ 0 & \text{other wise} \end{cases}, \quad (8)$$

where $w$ is the radius of each microlens, $d$ is the distance of the neighboring microlens, $i$ and $j$ denote the indices of microlenses in two perpendicular directions ($i = 0, j = 0$ for central microlenses). The phase transformation of the cross-type MLA can be written as:

$$\phi(\xi, \eta) = \sum_i \sum_j exp\left\{-j\frac{2\pi}{\lambda_{ex}}[(\xi-id)^2+(\eta-jd)^2]\right\}, and\ d > 2w, \quad (9)$$

where $\lambda_{ex}$ is the illumination wavelength. According to Eq. (8) and Eq. (9), we can obtain the Fraunhofer diffraction pattern at the focal plane of the MLA:

$$U_2(x,y) = \frac{exp\left[\frac{j\pi}{\lambda_{ex}f}(x^2+y^2)\right]}{j\lambda_{ex}f} \times \sum_i \sum_j \iint circ\left[\frac{\sqrt{(\xi-id)^2+(\eta-jd)^2}}{w}\right] \times exp\left\{-j\frac{2\pi}{\lambda_{ex}f}[(\xi-id)x+(\eta-jd)y]\right\} d(\xi-id)d(\eta-jd), \quad (10)$$

The integrand term in Eq. (10) is the Fourier transform of the input field at frequencies $k_{\xi-id} = \frac{x-id}{\lambda_{ex}f}$, $k_{\eta-jd} = \frac{y-jd}{\lambda_{ex}f}$,

$$U_2(x,y) = \frac{exp\left[\frac{j\pi}{\lambda_{ex}f}(x^2+y^2)\right]}{\lambda_{ex}f} \sum_i \sum_j \mathcal{F}\left\{circ\left[\frac{\sqrt{(\xi-id)^2+(\eta-jd)^2}}{w}\right]\right\}, \quad (11)$$

To calculate the Fourier transform of the $circ[\ ]$ function, we separate each microlens into unique polar coordinates and use the Hankel transform to obtain the Fourier transform of each microlens. Eq. (11) can be further derived as:

$$U_2(x,y) = \sum_i \sum_j \frac{exp\left[\frac{j\pi}{\lambda_{ex}f}(x^2+y^2)\right]}{j\lambda_{ex}f} \times w^2 \frac{J_1\left[\frac{2\pi w}{\lambda_{ex}f}\sqrt{(x-id)^2+(y-jd)^2}\right]}{\frac{w}{\lambda_{ex}f}\sqrt{(x-id)^2+(y-jd)^2}}, \quad (12)$$

where $J_1$ is the first-order Bessel function. For incoherent imaging, we only care about the distribution of the illumination intensity; thus, the intensity at the MLA focal plane is:

$$I_2(x,y) = |U_2(x,y)U_2^*(x,y)| = \sum_i \sum_j \left(\frac{w^2}{\lambda_{ex}f}\right)^2 \left\{\frac{J_1\left[\frac{2\pi w}{\lambda_{ex}f}\sqrt{(x-id)^2+(y-jd)^2}\right]}{\frac{w}{\lambda_{ex}f}\sqrt{(x-id)^2+(y-jd)^2}}\right\}^2, \quad (13)$$

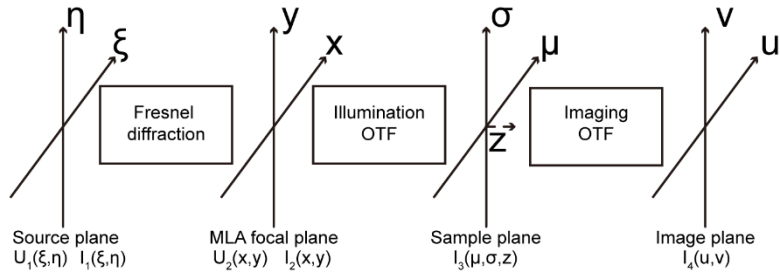

Fig. 2. Coordinate system of illumination and imaging process.

For the HiLo microscope, we use the incoherent imaging theory to conjugate this pattern to the sample plane for structured illumination.

We assume the blue dashed rectangular region in Fig. 1 is a telecentric configuration. The objective lens is used for illuminating and imaging the samples. According to the incoherent imaging theory, the intensity distribution on the sample plane is:

$$I_3(\mu,\sigma,z)|_{z=z_i} = \iint |h(\mu-x,\sigma-y)|_{z=z_i}|^2 \times I_{3-geometric}(\mu,\sigma)dxdy = \iint |h(\mu-x,\sigma-y)|_{z=z_i}|^2 \times I_2(Mx,My)dxdy, \tag{14}$$

where $h|_{z=z_i}$ is the coherent impulse response function for the defocus distance $z$, the coordinate $(\mu,\sigma)$ is related with $(\mu=Mx,\sigma=My)$, and $M$ is the lateral magnification. $z_i=0$ is the in-focus plane. $I_{3-geometric}(\mu,\sigma)$ is the scaled ideal geometric image copy of $I_2(x,y)$. Due to the telecentric configuration, $I_{3-geometric}(\mu,\sigma)$ remains the same for different $z$ distances. The corresponding spectrum of Eq. (14) can be expressed as:

$$G_3(f_\mu,f_\sigma)|_{z=z_i} = \mathcal{H}_{ill}(f_\mu,f_\sigma)|_{z=z_i} G_g(f_\mu,f_\sigma), \tag{15}$$

where $\mathcal{H}_{ill}(f_\mu,f_\sigma)|_{z=z_i}$ is the illumination optical transfer function (OTF) at a defocus distance $z_i$. Thus, Eq. (14) can be rewritten as:

$$I_3(\mu,\sigma,z)|_{z=z_i} = \mathcal{F}^{-1}\{\mathcal{H}_{ill}(f_\mu,f_\sigma)|_{z=z_i} G_g(f_\mu,f_\sigma)\}, \tag{16}$$

To get the defocused-OTF, we should know the corresponding coherent transfer function. Here, we assume the optical system is aberration-free, and the coherent transfer function of the system is:

$$\mathcal{H}_{ill}(f_\mu,f_\sigma) = circ\left(\frac{\lambda_{ex}z_{xp}\sqrt{f_\mu^2+f_\sigma^2}}{w_{xp}}\right)exp\left\{j\frac{2\pi}{\lambda_{ex}}W_d\left[\frac{(\lambda_{ex}z_{xp})^2(f_\mu^2+f_\sigma^2)}{w_{xp}^2}\right]\right\}, \tag{17}$$

where $z_{xp}$ and $w_{xp}$ are the exit pupil distance and exit pupil radius, respectively, $NA_{obj} \approx w_{xp}/z_{xp}$, $NA_{obj}$ is the numerical aperture of the objective lens, and $W_d$ is the Seidel defocus coefficient. Also, we have $W_d = \left(NA_{obj}^2/2\right) \cdot z_i$ [37]. Eq. (17) can be therefore rewritten as:

$$\mathcal{H}_{ill}(f_\mu,f_\sigma) = circ\left(\frac{\sqrt{f_\mu^2+f_\sigma^2}}{f_0}\right)exp\{j\pi z_i \lambda_{ex}(f_\mu^2+f_\sigma^2)\}, \tag{18}$$

where $f_0$ is $\lambda z_{xp}/w_{xp}$. The sign assignment of wavefront error is the same as [37]. For convenience, we recall it here. In Fig. 3, when the defocused wavefront converges to the right side of the ideal focus point and the corresponding defocused wavefront on the left side, $W_d$ and $z_i$ are positive.

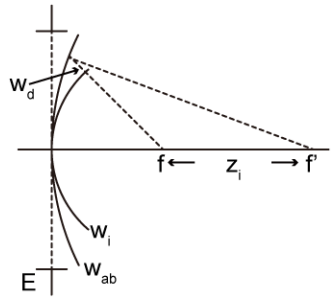

Fig. 3. Illustration of defocus. *E* is the exit pupil, wavefront $w_i$ converges to the focus point *f*, and the defocus wavefront $w_{ab}$ is centered on the axis at the defocus point *f'*.

According to the analytic expression for the OTF of a circular asymmetric optical system with defocused aberration, which was first given by Hopkins [38], the OTF of the HiLo illumination can be rewritten as:

$$\mathcal{H}_{ill}(\rho) = \frac{4}{\pi a}\cos\left(\frac{1}{2}\alpha\rho\right)\left\{\beta J_1(\alpha) + \sum_{n=1}^{\infty}(-1)^{n+1}\frac{\sin(2n\beta)}{2n}[J_{2n-1}(\alpha) - J_{2n+1}(\alpha)]\right\} \\ - \frac{4}{\pi a}\sin\left(\frac{1}{2}\alpha\rho\right)\sum_{n=1}^{\infty}(-1)^n\frac{\sin[(2n+1)\beta]}{2n+1}[J_{2n}(\alpha) - J_{2(n+1)}(\alpha)], \quad (19)$$

where $\alpha = \frac{2\pi w_{xp}^2 z_i}{\lambda_{ex} z_{xp}^2}\rho$, $\beta = \cos^{-1}\left(\frac{\rho}{2}\right)$, $\rho = \sqrt{s^2 + \tau^2}$, $s$ and $\tau$ are normalized spatial frequency components defined as $\lambda_{ex} z_{xp} f_\mu / w_{xp}$ and $\lambda_{ex} z_{xp} f_\sigma / w_{xp}$, respectively.

The high-order Bessel function makes Eq. (19) converge rather slowly. In simulations, we set $z_i = 0.5\lambda$ for each step, and the approximation of Eq. (19) can also obtain accurate results [39]. $\mathcal{H}_{ill}(f_\mu, f_\sigma)|_{z=z_i}$ in Eq. (16) is:

$$\mathcal{H}_{ill}(f_\mu, f_\sigma)|_{z=z_i} = \mathcal{H}_{ill}(\rho) = \begin{cases} 2(1 - 0.69\rho + 0.076\rho^2 + 0.043\rho^3)\left[\frac{J_1(\alpha - 0.5\alpha\rho)}{(\alpha - 0.5\alpha\rho)}\right], & |\rho| < 2 \\ 0, & |\rho| \geq 2 \end{cases}, \quad (20)$$

We calculated the Fourier transform of $I_2(Mx, My)$ to obtain $G_g(f_\mu, f_\sigma)$ in Eq. (16). Recalling Eq. (13) and using the autocorrelation theorem, the Fourier transform of $I_2(Mx, My)$ is:

$$\mathcal{F}\{I_2(Mx, My)\} = \mathcal{F}\{|U_2(Mx, My)|^2\} = \frac{1}{M^4}\left[ACF\left\langle G_2\left(\frac{f_x}{M}, \frac{f_y}{M}\right)\right\rangle\right], \quad (21)$$

where $ACF\langle\;\rangle$ is the correlation operator and $G_2\left(\frac{f_x}{M}, \frac{f_y}{M}\right)$ is the scaling spectrum of $U_2$. Using the convolution and successive transform theorems, neglecting the constant term in Eq. (11), we can obtain the following:

$$G_2\left(\frac{f_x}{M}, \frac{f_y}{M}\right) = \mathcal{F}\left\{\exp\left[\frac{j\pi M^2(x^2+y^2)}{\lambda_{ex}f}\right]\right\} \otimes \sum_i \sum_j circ\left[\frac{\sqrt{[M(x-id)]^2 + [M(y-jd)]^2}}{Mw}\right], \quad (22)$$

where $\otimes$ is the convolution operator. Combing Eqs (16), (20), (21), and (22), we can rewrite the 3D illumination pattern on the sample volume as:

$$I_3(\mu, \sigma, z)|_{z=z_i} = \mathcal{F}^{-1}\left\{\mathcal{H}_{ill}(\rho) \times ACF\left\langle \mathcal{F}\left\{\exp\left[\frac{j\pi M^2(x^2+y^2)}{\lambda_{ex}f}\right]\right\} \otimes \sum_i \sum_j circ\left[\frac{\sqrt{[M(x-id)]^2 + [M(y-jd)]^2}}{Mw}\right]\right\rangle\right\}, \quad (23)$$

From Eq. (7), we can rewrite it in the spectrum domain:

$$I_{4s}(u, v) = \sum_{z_i=0}^{n} \mathcal{F}^{-1}\{\mathcal{H}_{ima}(\rho)\mathcal{F}\{O(\mu, \sigma, z_i)I_3(\mu, \sigma, z)|_{z=z_i}\}\}, \quad (24)$$

where $\mathcal{H}_{ima}(\rho)$ is the same as Eq. (20), except that the emission wavelength $\lambda_{em}$ substitutes the illumination wavelength $\lambda_{ex}$.

## 3. Simulations

### 3.1 3D illumination pattern on the sample volume

We simulated different 3D illumination patterns on the sample volume. The illumination wavelength is 520nm, the illumination NA is 0.5, and the sample volume is 200*200*100μm (width*length*height). We set MLA's NA from 0.006 to 0.01 with a step of 0.001. Each microlens is 50μm in diameter; the microlens pitch (distance between adjacent microlenses) is from 60μm to 140μm with a 20μm step. Fig. 4 shows several normalized 3D illumination patterns created by different MLAs. The x-y section shows the intensity distribution on the focus plane ($z_i = 0$μm), and the x-z section shows the axial intensity distribution of the central slice (y = 0) from $z_i = 0$μm to $z_i = 100$μm. The magnification of the illumination light path is 0.1; therefore, the pattern period on the sample plane in Fig. 4 is ten times smaller than MLA's lens pitch (6μm, 10μm, and 14μm).

When using MLA to create SI on the sample, its microlens pitch and NA determine the modulation depth (MD). Fig. 5 shows horizontal line profiles of the focus plane (x-y section, y = 0 in Fig. 4). In the same MLA type, a higher microlens pitch increases MD. Beyond that, increasing NA also causes a higher MD. According to Eqs (8) and (18), the intensity distribution formed by each microlens at the sample plane is an Airy pattern. Therefore, if the period (corresponding to MLA's microlens pitch) and NA are too large, the side lobes of each focus point will sum up and cause oscillation (purple arrows in Fig. 5). We define MD as $MD(\%) = (I_{max} - I_{min})/(I_{max} + I_{min})$, where $I_{max}$ and $I_{min}$ are the profile's maximum and minimum intensities. Fig. 6 shows relationships between MD, NA, and the period for different MLAs. As expected, increasing the period and NA can obtain higher MD rates. Because of the structure's similarity, illumination patterns formed by the cross-type and hexagon-type MLAs have similar MD rates and trends. When the period and the NA are both low, cylinder-type MLAs can create patterns with higher MD rates ($> 50\%$).

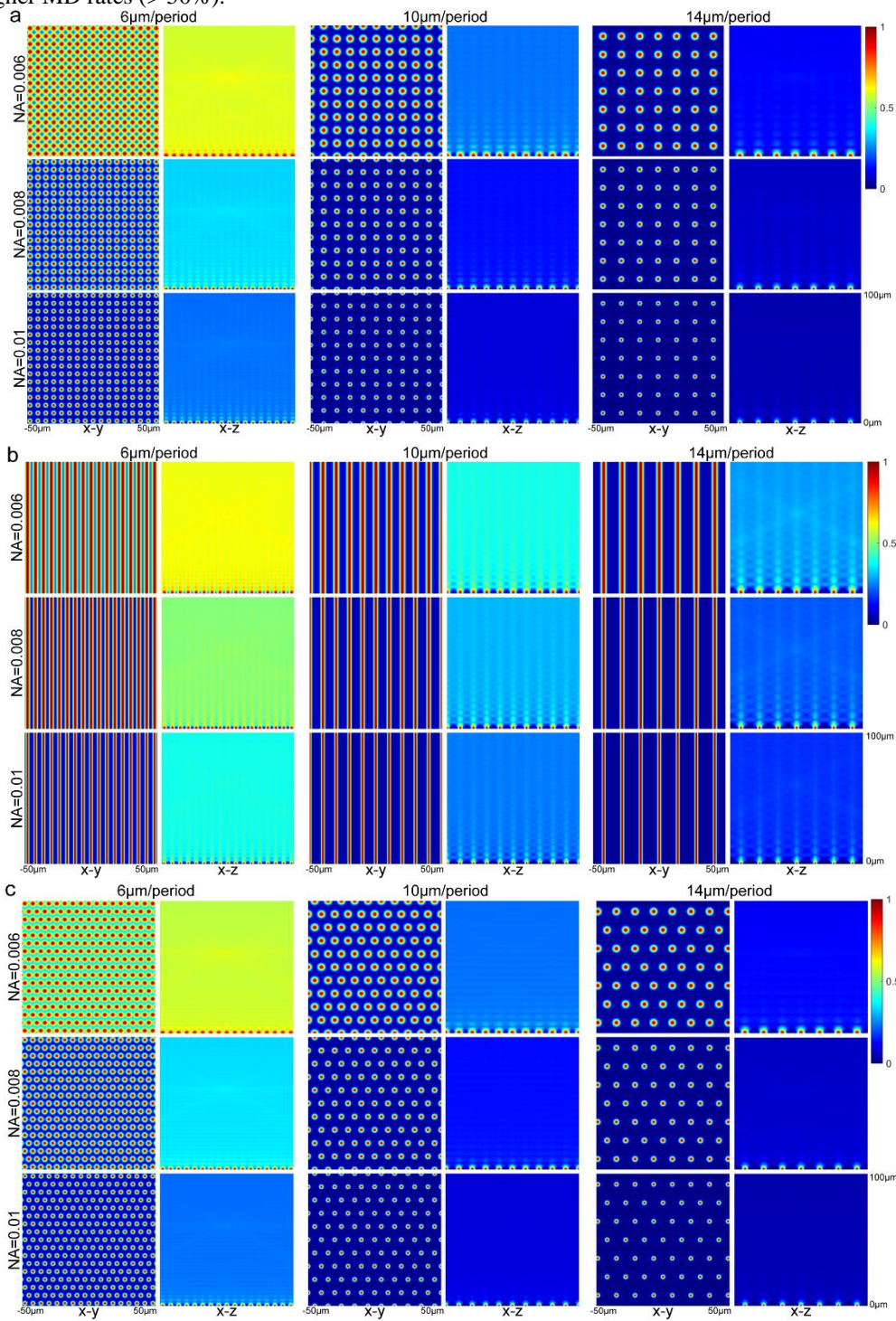

Fig. 4. 3D illumination intensity distribution on the sample volume created by cross-type (a), cylinder-type (b), and hexagon-type (c) MLAs. Each picture is self-normalized, and the NA is MLA's numerical aperture. The period is the distance between adjacent focus points on the x-y focus plane, which MLA creates.

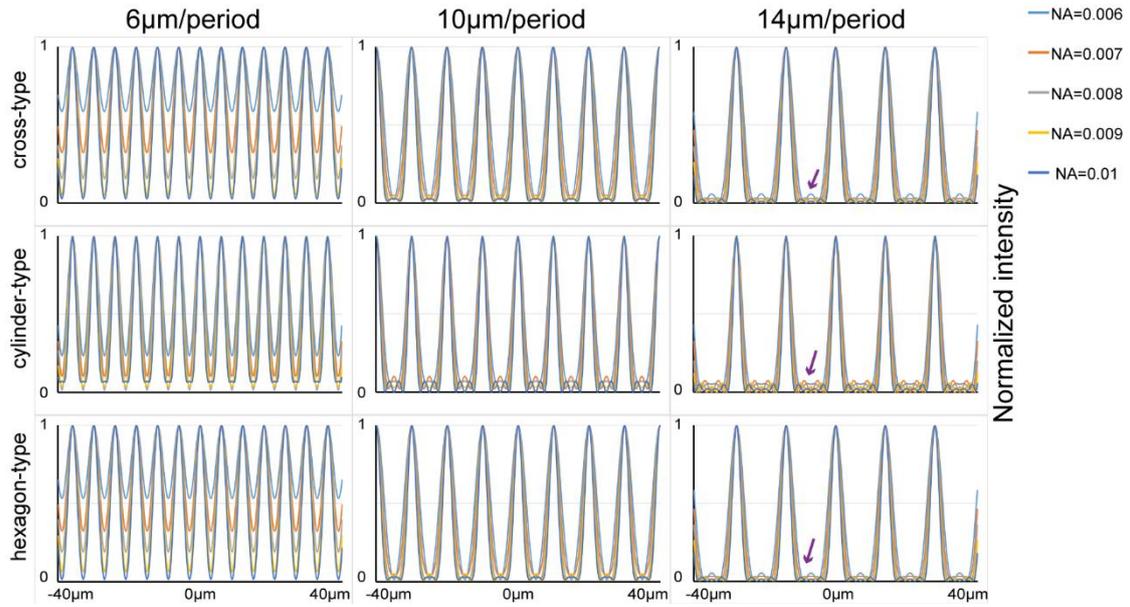

Fig. 5. Horizontal line profiles of illumination patterns on the in-focus plane (x-y section, y=0). Different color lines represent microlens NA.

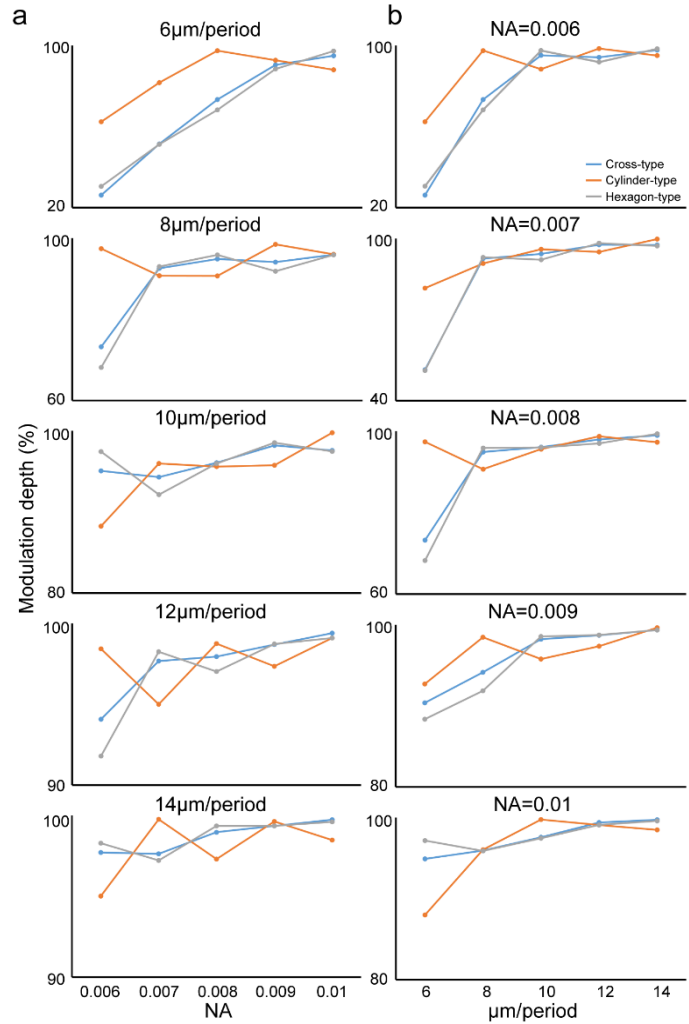

Fig. 6. Relationships between MD and microlens NA (a), MD and pattern period (b). Different color lines represent MLA's type. Scales are different for better visualization.

*3.2 The axial resolution of illumination patterns on the sample volume*

The relationship between MLA and axial resolution of illumination patterns on the sample volume is also numerically studied. Fig. 7 shows the central (x = y = 0) axial line profiles of illumination patterns on the sample volume. The axial range is 0μm to 10μm. Purple arrows denote the first global minimal value along the axial line. We denoted this position as $z_{min}$, whose corresponding intensity is $I_{z-min}$.

Similarly, the in-focus position $z_{max}$ has the maximum intensity $I_{z-max}$. $\delta I$ is $I_{z-max} - I_{z-min}$. To quantify the axial resolution, we defined the axial FWHM as $2z_{FWHM}$. $z_{FWHM}$ is the axial position when the axial intensity is half the difference between maximum and first side lobe minimum intensity. For clearance, these definitions are annotated in Fig. 7, center picture. Fig. 7 shows that Cross-type and Hexagon-type line profiles are similar, and Cylinder-type lines are always higher than the others. As expected, a higher NA increases $\delta I$. Beyond that, $z_{min}$ seems only determined by the period of the illumination pattern, and a higher period would bring a larger $z_{min}$. We did a quantitative analysis to discover how the MLA type affects the axial resolution of illumination patterns. Table 1. summarizes how $z_{min}$ behaves in terms of the MLA type, the periods, and NA. Obviously, $z_{min}$ depends only on the period of the illumination pattern.

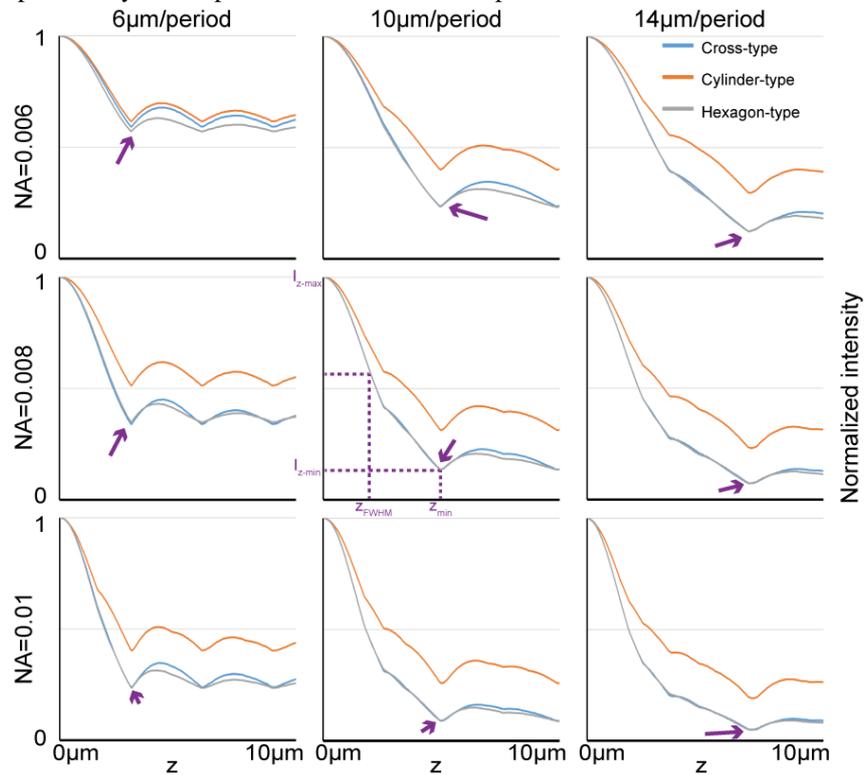

Fig.7. Axial line profiles of illumination patterns at x = y = 0. The purple arrow denotes the first global minimal position along the axial line. The central picture shows detailed positions about $z_{min}$, $z_{FWHM}$, $I_{z-min}$, and $I_{z-max}$.

**Table 1. The value of $z_{min}$ (μm as the unit) in terms of the MLA type, the period, and NA. Cro, cyl, and hex denote cross-type, cylinder-type, and hexagon-type MLAs.**

| Period  NA | 6μm/period | | | 8μm/period | | | 10μm/period | | | 12μm/period | | | 14μm/period | | |
|---|---|---|---|---|---|---|---|---|---|---|---|---|---|---|---|
| | cro | cyl | hex | cro | cyl | hex | cro | cyl | hex | cro | cyl | hex | cro | cyl | hex |
| 0.006 | 3 | 3 | 3 | 4 | 4 | 4 | 4.95 | 4.95 | 4.95 | 5.85 | 5.85 | 5.85 | 6.8 | 6.9 | 6.8 |
| 0.007 | 3 | 3 | 3 | 4 | 4 | 4 | 4.95 | 4.95 | 4.95 | 5.85 | 5.95 | 5.85 | 6.9 | 6.9 | 6.8 |
| 0.008 | 3 | 3 | 3 | 4 | 4 | 4 | 4.95 | 4.95 | 4.95 | 5.95 | 5.95 | 5.85 | 6.9 | 6.95 | 6.9 |
| 0.009 | 3 | 3 | 3 | 4 | 4 | 4 | 4.95 | 4.95 | 4.95 | 5.95 | 6 | 5.95 | 6.95 | 6.95 | 6.9 |
| 0.01 | 3 | 3 | 3 | 4 | 4 | 4 | 4.95 | 5.05 | 4.95 | 5.95 | 6 | 5.95 | 6.95 | 7 | 6.95 |

Fig. 8 shows the axial resolution ($z_{FWHM}$) changing with different MLA's NA and illumination pattern periods. The higher the MLA's NA or, the lower the illumination pattern period, the better the axial resolution. In Fig. 8(a), the gradient is steeper with a higher period. On the other hand, the gradient is gentle with higher NA in Fig. 8(b). In addition, we found that the cylinder-type MLA has worse axial resolution when compared to the others. Because the cylinder-type MLA creates periodic features on the focus plane only in a one-dimensional direction. The intensity variation only exists in one direction; however, in cross-type and hexagon-type, this variation is in two orthogonal directions, and the propagating light field changes more quickly. Therefore, $z_{FWHM}$ in cross- and hexagon-types are smaller than $z_{FWHM}$ in the cylindrical type.

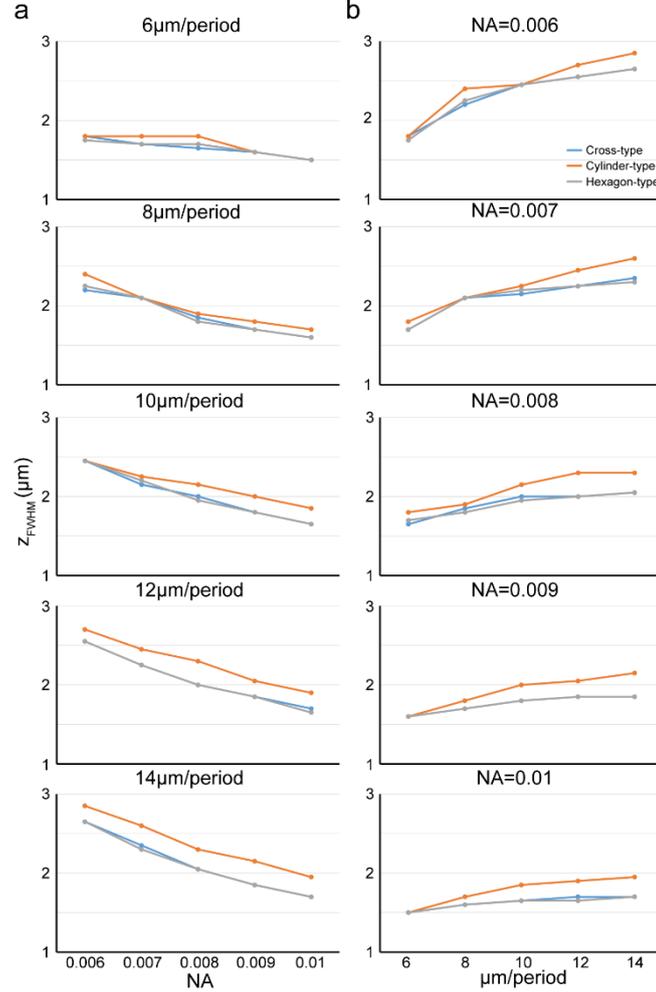

Fig.8. Relationships between axial resolution ($z_{FWHM}$) and microlens NA (a), axial resolution ($z_{FWHM}$) and pattern period (b). Different color lines represent MLA's type.

## *3.3 HiLo imaging simulation*

To understand the effects of MLA-generated SI patterns on the HiLo image, we modeled a fluorescent block for simulating incoherent widefield and HiLo imaging (Fig. 9(a) and (b)). According to Eq. (6), we first multiplied the sample with a specific 3D illumination pattern. Then, we used the incoherent imaging theory to acquire SI images on the camera plane. Similarly, we used the same method to obtain widefield images (Fig. 9(c)), except that the fluorescent block was multiplied by the all-one matrix instead of the MLA-generated SI pattern. In the simulations, we set the illumination and the imaging NA to 0.5, and the excitation and emission wavelengths are 520nm and 580nm, respectively. The magnification in the illumination light path is 10, and 1 in the imaging light path.

Fig. 10 shows images of the fluorescent block on the camera plane. SI columns are images under SI illumination, and HiLo columns show final HiLo pictures. The widefield image used as input for HiLo algorithms is shown in Fig. 9(c). In Fig. 10(a), the fluorescent target is excited by a cross-type illumination pattern, and others are excited by cylinder-type (Fig.10(b)) and hexagon-type (Fig.10(c)). When the pattern's period is 14μm/period, we found uneven artifacts on these HiLo images (white arrows). Especially when the pattern is the hexagon type (Fig.10(c)), artifacts seem the worst. The cross-type pattern also generates recognizable ununiform intensity (Fig.10(a)). In cylinder-type patterns, these imperfections are the mildest. Artifacts are caused by side lobes that sum up and cause oscillations (Purple arrows in Fig. 5). With larger lens pitch and higher NA, more side lobes will exist between each microlens, and the oscillation will become more serious. These artifacts are the mildest in cylinder-type patterns because side lobes and oscillation only exist in one dimension but two dimensions in both cross-type and hexagon-type. Furthermore, cross-type has fewer artifacts when compared with hexagon-type since its two-dimensional patterns are orthogonal and can be decomposed into *x* and *y* directions. However, in hexagon-type, these artifacts are correlated. To quantitively analyze how MLA-generated illumination patterns affect HiLo images, we traced line profiles on each HiLo image (Red line in Fig. 10(a). Same tracing positions in all HiLo images.).

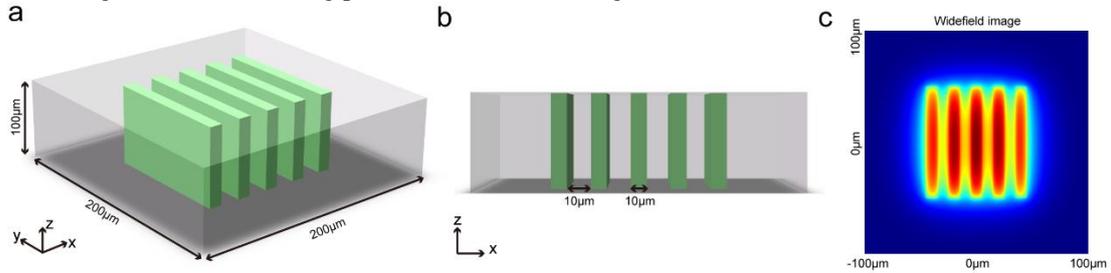

Fig. 9. Simulated fluorescent block for HiLo imaging (a and b). The green parts can be excited by 520nm wavelength illumination. The residual grey transparent parts cannot create a fluorescent signal. The emission wavelength is 580nm. (c) The self-normalized widefield image on the camera plane.

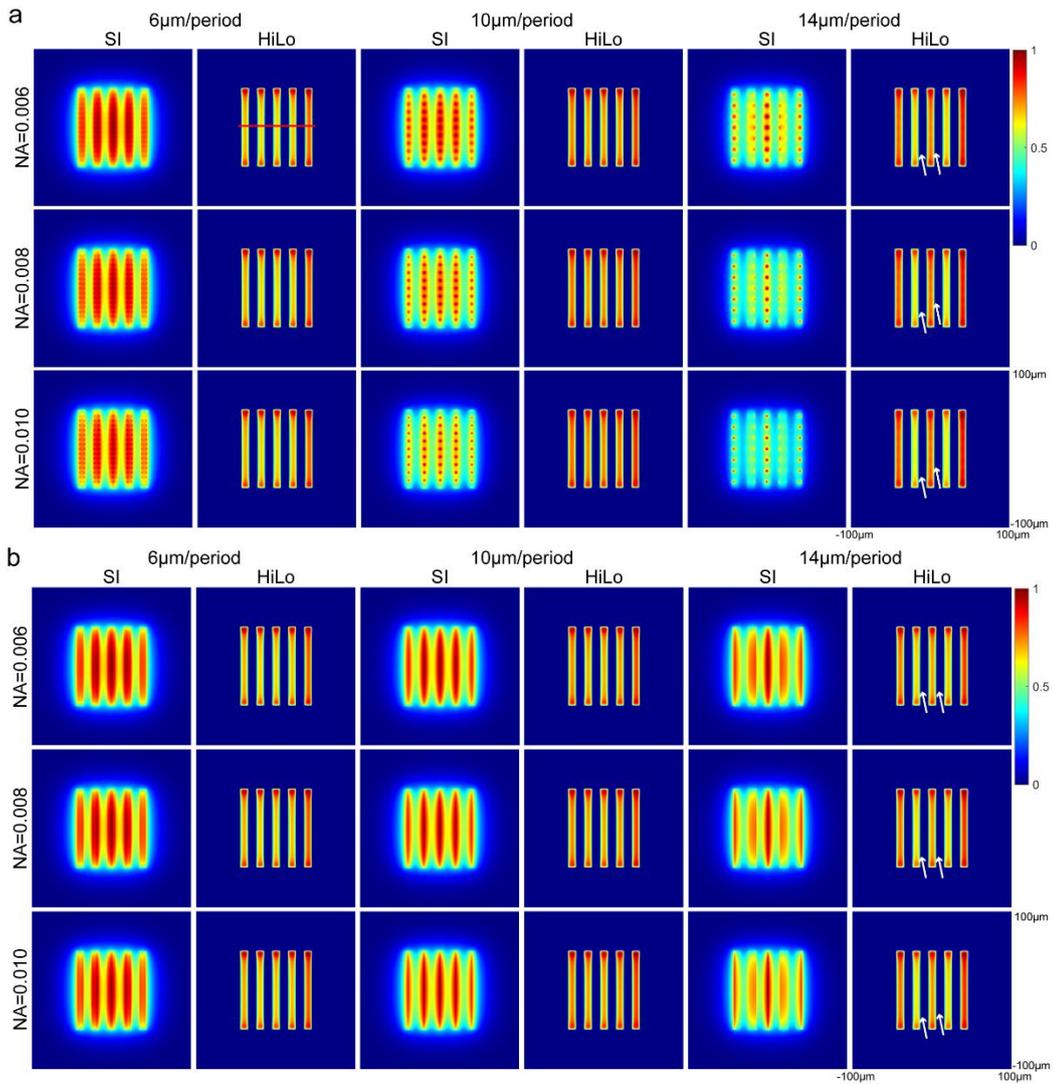

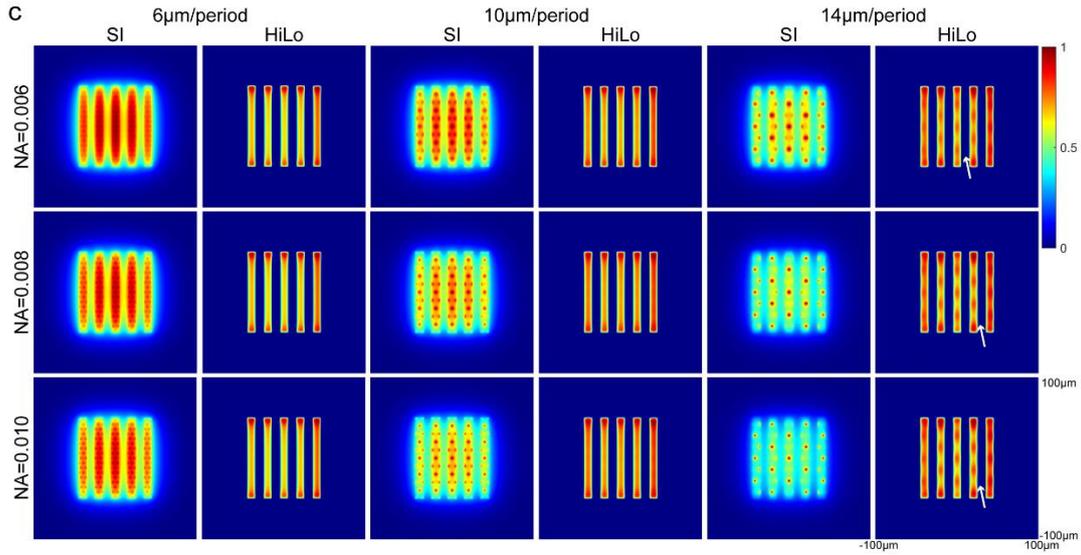

Fig. 10. Captured images on the camera plane. The corresponding widefield image is shown in Fig. 9(c). SI columns show images under different SI illumination. HiLo columns are the final HiLo images. Headlines are periods of corresponding illumination patterns. 8 μm/period and 12 μm/period are not shown. Different NAs are corresponding to MLA's numerical aperture. NA = 0.007 and NA = 0.009 are not shown. (a) Cross-type illumination pattern. (b) Cylinder-type. (c) Hexagon-type. Each image is normalized. White arrows depict imperfections.

In Fig.10, these images are similar; therefore, captured images of 8μm/period, 12μm/period, NA=0.007, and NA=0.009 are not shown. However, we traced line profiles on every condition, NA from 0.006 to 0.01 with a step of 0.001, and period from 6μm/period to 14μm/period with 2μm/period step. According to the line profiles shown in Fig. 11, the NA of MLA does not seem to affect the quality of the final HiLo image. Fig.11(a) (12μm/period column) shows a minor contrast enhancement with a higher NA. However, some artifacts are created if the period is too high (14μm/period). In Fig. 11(a), the contrast of the second and fourth lobes (purple arrows) is lower than others, and their trends become inverse (the higher the NA, the lower the contrast). In Fig.11(b), we can see clearly that a higher period can enhance the final HiLo image contrast, and this tendency is more evident for the cross-type illumination. Like Fig.11(a), artifacts appear when the period is 14μm/period.

Furthermore, we recorded the normalized local maximum intensity of the middle lobe in Fig.11. Table 2 shows how the intensity changes with different periods and NA. The period affects the intensity more significantly than MLA's NA. The best parameter setting is the cross-type pattern with 0.01NA and 12μm/period.

**Table 2. The normalized local maximum intensity of the middle lobe in Fig. 11. Abbreviations are the same in Table 1**

| Period NA | 6μm/period | | | 8μm/period | | | 10μm/period | | | 12μm/period | | | 14μm/period | | |
|---|---|---|---|---|---|---|---|---|---|---|---|---|---|---|---|
| | cro | cyl | hex | cro | cyl | hex | cro | cyl | hex | cro | cyl | hex | cro | cyl | hex |
| 0.006 | 0.84 | 0.844 | 0.838 | 0.85 | 0.852 | 0.85 | 0.863 | 0.862 | 0.862 | 0.916 | 0.868 | 0.888 | 0.873 | 0.859 | 0.867 |
| 0.007 | 0.842 | 0.846 | 0.841 | 0.853 | 0.855 | 0.853 | 0.866 | 0.864 | 0.866 | 0.928 | 0.872 | 0.895 | 0.875 | 0.861 | 0.868 |
| 0.008 | 0.844 | 0.848 | 0.843 | 0.855 | 0.856 | 0.856 | 0.869 | 0.865 | 0.869 | 0.936 | 0.875 | 0.898 | 0.877 | 0.861 | 0.869 |
| 0.009 | 0.845 | 0.85 | 0.844 | 0.857 | 0.857 | 0.859 | 0.871 | 0.867 | 0.871 | 0.942 | 0.877 | 0.9 | 0.878 | 0.862 | 0.869 |
| 0.01 | 0.847 | 0.852 | 0.846 | 0.858 | 0.858 | 0.86 | 0.872 | 0.868 | 0.873 | 0.946 | 0.879 | 0.901 | 0.879 | 0.863 | 0.87 |

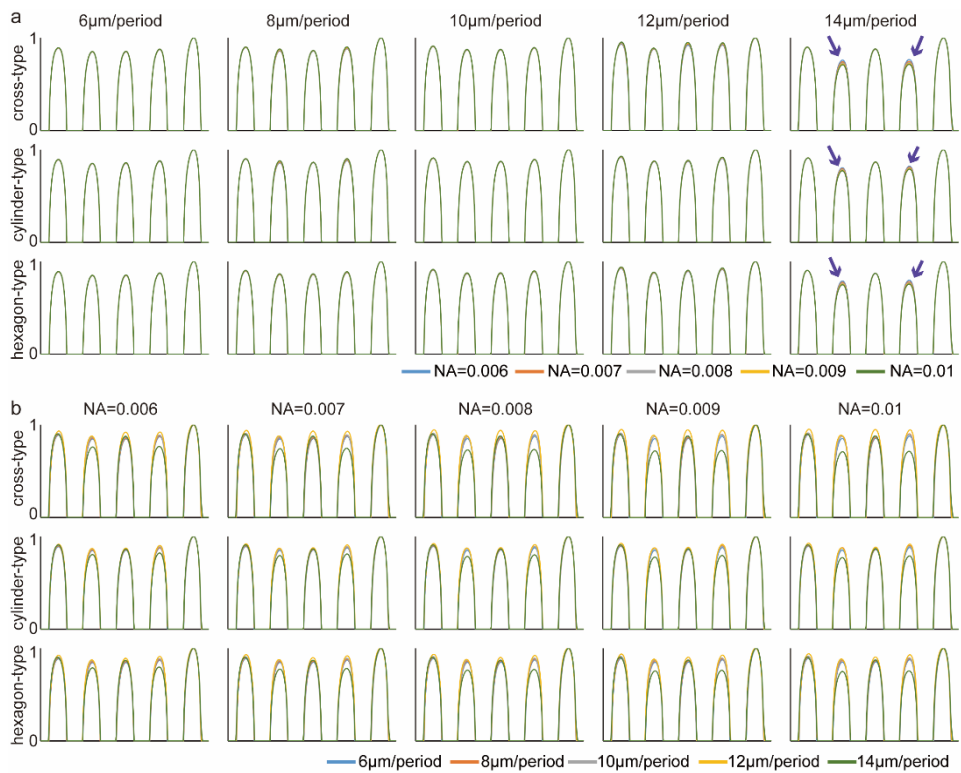

Fig. 11. Line profiles (along the red line in Fig.10) of HiLo images under different illumination patterns. In (a), we can see that the NA of MLA does not significantly affect the image quality. Interestingly, when the period is 14μm/period, the contrast of the second and fourth lobes (counted from left to right) is lower than others. (Depicted by purple arrows). The same phenomenon can be found in (b). In (b), the image contrast becomes better with a higher period, except 14μm/period

## 4. Discussion and prospect

   HiLo microscopy is a powerful and straightforward tool for optical sectioning. In most cases, there are two ways to generate SI patterns: Using a coherent light source and a rotated scattering glass to create speckle patterns or using an incoherent light source and DMD to generate customized periodic patterns. Nevertheless, the coherent light source and DMD make the system costly and bulky. Here, we found that using an incoherent light source and MLA can also realize HiLo microscopy, and this configuration can reduce HiLo's cost and system complexity.

   In this study, we choose three different MLA types (cross-, cylinder-, and hexagon-types) with different physical parameters (NA and the microlens pitch size) as diffractive components to generate SI patterns. Firstly, we used the Fresnel diffraction theory to study how different MLAs affect SI patterns. The modulation depth increases with a higher NA when the microlens pitch is small enough. Similarly, the modulation depth increases significantly with a broader lens pitch in low NA situations. However, when NA (the microlens pitch) is too high, the modulation depth would not be affected by the microlens pitch (NA).

   NA and the microlens pitch size need to be carefully chosen. A higher NA would minimize focus point size, and a higher period can lead these focus points side lobes to accumulate between each microlens, causing oscillations (Fig.5 purple arrows). These oscillations will create artifacts on the final HiLo images (Fig.10 red arrows). Here, we chose NA from 0.006 to 0.01 to prevent the point size is too small. In terms of other MLAs type, the cross-type and hexagon-type perform similarly, and the cylinder-type MLA can generate SI patterns with a better modulation depth when NA or the microlens pitch is low. We also discussed the axial resolution of SI patterns. As expected, when the microlens pitch lens is unavailable, a higher NA can deliver a better axial resolution. This trend is evident with a bigger microlens pitch. Conversely, the axial resolution becomes worse with a higher microlens pitch when NA is unaltered, and this phenomenon is more pronounced in low NA conditions. When MLA's parameters are the same, the cylinder-type MLA has the worst axial resolution, and the cross-type and the hexagon-type MLA perform similarly.

   Continuously, we multiplied the simulation-generated fluorescent block with SI patterns, using the incoherent imaging theory and HiLo algorithms to investigate the links between HiLo images and MLAs. We found subtle differences among these HiLo images with different MLA's NA, and a higher NA can enhance the image contrast negligibly. Regarding the microlens pitch, a higher distance leads to better image contrast. Interestingly, a large microlens pitch can introduce artifacts and ununiform intensity to final HiLo images. We also found that the hexagon-type MLA is the most susceptible to artifacts. Using the cross-type MLA with NA of 0.01 and the microlens pitch of 120 μm to generate SI patterns can obtain the best HiLo images. In cylinder-type and hexagon-type, the best parameter is NA = 0.01 with a 120μm microlens pitch. Besides, due to the simplicity of corresponding periodic patterns, less oscillation can be found in cylinder-type-generated illumination patterns. We can obtain HiLo images with the mildest artifacts using cylinder-type MLA.

Overall, in this study, we found that the cylinder-type MLA can be used for HiLo imaging with the mildest artifacts due to the simplicity of the corresponding structure. However, this aspect also leads to its corresponding light intensity field only varying in one dimension, which minimizes the axial resolution. Cross-type and hexagon-type MLAs can obtain a better axial resolution but tend to produce more artifacts. Therefore, there exists a trade-off between the artifacts-free and better axial resolution. Researchers should carefully choose the appropriate MLA for HiLo imaging.

This study finds a new method for HiLo imaging. During the simulation, we found several interesting phenomena, which will be further investigated through theoretical analysis and mathematical deduction in future. These include:

- A quantitative link between the microlens pitch and NA (e.g., ratio).
- Theoretical and mathematical analysis about why the $z_{min}$ in Fig.7 is unrelated to NA.
- If translating different distances of each line of the microlens in cross-type MLA, how this affects the final HiLo image?

## 5. Conclusion

In summary, we found that MLA is practical and cost-effective for generating structured illumination for HiLo microscopy. Our study shows how MLA's parameters (MLA's type, NA, and microlens pitch) affect final HiLo images. Simultaneously, we found the best MLA parameters in our study. To our knowledge, this is the first study about using proper MLAs to realize HiLo microscopy. A more detailed analysis will be conducted for further investigation, including a more comprehensive NA range, a larger microlens pitch, the ratio between NA and the lens pitch, and multiple microlens arrangements. Furthermore, we believe this study can guide more researchers to establish MLA-based HiLo microscope systems. Soon, such tools could help researchers acquire good-quality biomedical images and analyze them efficiently and cheaply.

## Acknowledgements

The research has been supported by the Engineering and Physical Sciences Research Council under EPSRC Grant: EP/T00097X/1, the Royal Society of Edinburgh, and the China Scholarship Council.